# An Asymmetric Capital Asset Pricing Model

Abdulnasser Hatemi-J

Department of Economics and Finance, UAE University

Email: AHatemi@uaue.ac.ae

Abstract

Providing a measure of market risk is an important issue for investors and financial institutions. However, the existing models for this purpose are per definition symmetric. The current paper introduces an asymmetric capital asset pricing model for measurement of the market risk. It explicitly accounts for the fact that falling prices determine the risk for a long position in the risky asset and the rising prices govern the risk for a short position. Thus, a position dependent market risk measure that is provided accords better with reality. The empirical application reveals that Apple stock is more volatile than the market only for the short seller. Surprisingly, the investor that has a long position in this stock is facing a lower volatility than the market. This property is not captured by the standard asset pricing model, which has important implications for the expected returns and hedging designs.

**Keywords**: Market Risk, Company Risk, Long Position, Short Position, Apple, Nasdaq.
**JEL Classifications**: C22, C12, G12.

1. **Introduction**

Determining the fair rate of return for a risky asset such as a stock is of fundamental importance in computational finance. This risk consists of two components—namely (1) systematic or market risk and (2) unsystematic or firm specific risk. The firm specific risk can be neutralized via efficient portfolio diversification via methods suggested by Markowitz (1952) or Hatemi-J and El-Khatib (2015) and Hatemi-J, Hajji and El-Khatib (2022). The market risk is not diversifiable, however. An alternative approach is needed for hedging the market risk. The capital asset pricing model (CAPM) developed by Sharpe (1964) is considered as the first attempt to provide a measure for market risk, which can also function as the optimal hedge ratio for neutralizing this risk via futures. Even though the CAPM is attributed to Sharpe (1964), according to literature Treynor (1961), Littner (1965a, b) and Mossin (1966) also deserve to be credited for development of this model independently. For a review of the CAPM see Fama and French (2004). For new extensions of this model see Habis (2024). Also, for the advantages of estimating the asset pricing model via neural networks method see Chen et. al., (2024).[1] The fundamental implicit assumption by Sharpe (1964) is that the investor is has a long position in the risky asset. However, it is also possible that the investor undertakes a short position in the underlying risky asset. This paper suggests an asymmetric version of the CAPM that explicitly takes the position of the trader into account. This is an important issue because the source of risk for the long position is falling prices while the source of risk for a short seller is the rising prices. The standard CAPM does not make a separation of the direction of the price changes and any price change is part of the risk regardless of if it is positive or negative.

The rest of this paper is structured as follows. In the next section symmetric and asymmetric CAPMs are presented. In section 3 a numerical application is provided. Conclusions are presented at the end.

## 2. Symmetric and Asymmetric CAPMs

In the first part of this section, we describe the standard CAPM, which we call the symmetric CAPM. In the second subsection we introduce the asymmetric version of this model.

### 2.1 The Symmetric CAPM

According to the symmetric model the unsystematic (i.e., firm specific) risk can be neutralized via portfolio diversification. Thus, in a market that is at equilibrium the investors will be only compensated for the systematic (i.e., the market) risk.[2] The standard CAPM is based on the following equation:

$$R_i = R_F + \beta_i(R_M - R_F) \qquad (1)^3$$

---

[1] The CAPM model fails empirically in some cases. Andrei et. al., (2023) provide an explanation for this failure. For critical review of the literature on asset pricing via machine learning approach see Bagnara M. (2024). For alternative stock pricing models see, among others, Gordon (1962), Gordon and Gould (1978), Gordon and Shapiro (1956), and Hatemi-J and El-Khatib (2022).
[2] An example of systematic risk (market risk) is the change in the interest rate that affects all the firms. A change in the organizational form of a firm is an example of unsystematic risk (firm specific risk) because this change will only affect that firm and not all other firms in the market.
[3] It should be mentioned that equation (1) can be easily estimated by an ordinary least squares regression.

Where $R_i$ is the rate of return on risky security $i$, $R_F$ is the risk-free return, and $R_M$ is return on the market. Note that $(R_M - R_F)$ is the market risk premium and $\beta_i(R_M - R_F)$ is the risk premium of security $i$. The risk premium of a risky asset is defined as the additional expected return than the risk-free rate that the investors require to invest in that asset. The denotation $\beta_i$ represents the beta of security $i$, and it can be calculated by using the following equation:

$$\beta_i = \frac{Cov(R_i, R_M)}{\sigma^2(R_M)}$$

Where $Cov(R_i, R_M)$ is the covariance between the return of security $i$ and the return of the market. The denotation $\sigma^2(R_M)$ is the variance of the market return. Beta is a measure of market risk. If the beta of a security is less than one it means that that security is less risky than the market. Likewise, if the beta of a security is higher than one it means that that security is riskier than the market. Note that the beta of the market is equal to one per definition. Thus, if a security has a beta equal to one, that security is exactly as risky as the market. According to the CAPM a security that has a beta equal to zero can be considered as risk-free.

The CAPM has several important advantages in financial markets. It simplifies the calculation of inputs that are necessary for establishing a portfolio. This is especially the case if the number of securities considered for including into the portfolio is large. Another advantage of the CAPM is that it provides a measure of the required rate of return that is theoretically appropriate and can be used as a fair cost of capital. This is an important and useful issue in making appropriate capital budget decisions. The CAPM can also be used to determine whether a security is underpriced, overpriced, or correctly priced.[4]

## 2.2 The Asymmetric CAPM

The asymmetric version of this model provides a position dependent measure of risk depending on whether the trader is owner of the stock or is short selling. For the owner of the stock the risk is defined by the falling prices, thus, a logical conclusion is to base the risk on negative returns only. That is, the following model can measure the market risk for the long position of the stock (or a portfolio of stocks):

---

[4] Note that the pricing of an initial public offering (IPO) below its market value is an example of underpricing. When the offer price is lower than the price of the first trade, the stock is underpriced.

$$R_i^- = R_F + \beta_i^-(R_M^- - R_F) \tag{2}$$

Where the negative components are defined as $R_i^- := min(R_i, 0)$ and $R_M^- := min(R_M, 0)$. Note that since the risk free rate is assumed to be constant it is not going to impact the beta measure. The value of $\beta_i^-$ is estimated as the following:

$$\beta_i^- = \frac{Cov(R_i^-, R_M^-)}{\sigma^2(R_M^-)}$$

here $Cov(R_i^-, R_M^-)$ is the covariance between the negative returns of security $i$ and the negative returns of the market. The denotation $\sigma^2(R_M^-)$ is the variance of the negative market returns. An investor with a long position in the underlying asset is concerned with $\beta_i^-$ as a measure of market risk since the source of risk is the falling price in this case while increasing price is just good.

Similarly, the source of risk for a short seller is rising prices, positive returns should be used for measuring the market risk. Thus, the following model can be used for measuring the market risk for a short position in the asset:

$$R_i^+ = R_F + \beta_i^+(R_M^+ - R_F) \tag{3}$$

Where the positive components are defined as $R_i^+ := max(R_i, 0)$ and $R_M^+ := max(R_M, 0)$ and $\beta_i^+$ is estimated numerically by the following ratio:

$$\beta_i^+ = \frac{Cov(R_i^+, R_M^+)}{\sigma^2(R_M^+)}$$

$Cov(R_i^+, R_M^+)$ represents the covariance between the positive returns of security $i$ and the positive returns of the market. The denotation $\sigma^2(R_M^+)$ signifies the variance of the positive market returns. An investor with a short position in the underlying asset is concerned with $\beta_i^+$ as a measure of market risk. This is the case because the source of risk for a short seller is the rising price whereas decreasing price is the good news.
.

It is also operational to estimate each model by making use of position dependent measures of time-varying risk such as the method suggested by Hatemi-J (2013, 2023). Additional potential extension is providing an asymmetric version of the multivariate factor asset pricing model based the arbitrage pricing theory introduced by Ross (1976).

## 3. Application

A numerical example is provided for measuring the market risk for Apple stock during the period of September 2020 until March 2024 on monthly basis. The choice of the sample period is based on covering the period after a four to one stock split that took place on 31 of August 2020. The NASDAQ Composite is used as a measure of the market index during the period. The source of the data is Yahoo Finance. The empirical findings are presented in Table 1.

Table 1: The Estimation Results.

| Parameters | Estimated Values | P-values |
|---|---|---|
| $\beta_i$ | 1.027776 | <0.00000 |
| $\beta_i^+$ | 1.007963 | <0.00000 |
| $\beta_i^-$ | 0.821638 | <0.00000 |

As it is evident from the empirical results, the standard beta measure (i.e., the symmetric one) is equal to 1.027776. Thus, a trader that has a long position in the stock needs to have 1.027776 units of short futures for each unit of spot to hedge. On the other hand, the trader with a short position in the stock needs to have 1.027776 units of long futures for each unit of spot to hedge. However, the results change to some extent if asymmetric beta measures are used. In this case, an investor with a short position in the stock needs to have 1.007963 unit of long futures for optimal hedging. While an investor with a long position in the stock should have 0.821638 units of short futures for optimal hedging. Interestingly, this asymmetric approach reveals that Apple stock is riskier than the market if and only if the investor has a short position in the stock. An investor with a long position in the Apple stock faces a lower risk than the market since its beta is lower than one. If the position dependent measures of market risk were not used one would conclude that the investor of Apple stock is facing higher risk than the market regardless of if the investor is the owner of the stock or is short selling. It should be mentioned that the estimated parameter is statistically significant in each case.

To test desirable statistical assumptions for a good model based on the least squares method that is used to estimate the beta values, a number of diagnostic tests are conducted. The empirical findings of these tests are presented in Table 2. The results reveal that the null hypothesis that each assumption is fulfilled cannot be rejected at the five percent significance level for each model.

Table 2: The Diagnostic Test Results.

| Diagnostic Tests | The Model for $\beta_i$ | The Model for $\beta_i^+$ | The Model for $\beta_i^-$ |
|---|---|---|---|
| Jarque-Bera | 0.4952 | 0.1681 | 0.0886 |
| Breusch-Godfrey | 0.3588 | 0.9756 | 0.6422 |
| Breusch-Godfrey-Pagan | 0.6581 | 0.2247 | 0.0661 |

Notes. Jarque-Bera tests for the null hypothesis of normality. Breusch-Godfrey tests for the null hypothesis of no serial correlation. Breusch-Godfrey-Pagan tests for the null hypothesis of Homoscedasticity. The p-values are presented.

## 4. Concluding Remarks

It is a fact that the source of risk for a long position in a risky asset is determined by the falling prices while the source of the risk for a short position in the underlying asset is the rising prices. The current paper argues that explicitly considering this asymmetric property is important for successful risk management and hedging purposes. A position dependent measure of market risk is presented in this paper. A numerical example is also provided, which reveals that an investor of the Apple stock faces a higher level of risk compared to the market for a short position in that stock only. However, for a long position in the Apple stock, the risk of this stock is lower than the market risk. These empirical findings based on the suggested approach have important implications in terms of the expected returns and for designing optimal hedging strategies in each case. Future applications of this method will demonstrate, however, whether this asymmetric property of the market risk is a general phenomenon for different classes of risk assets or not.

**Acknowledgements**

The current paper is partially funded by the CBE Annual Research Program (CARP) 2024 granted by the United Arab Emirates University, which is immensely appreciated. The usual disclaimer applies, however.